\documentclass[prl,twocolumn,superscriptaddress,showpacs]{revtex4}

\usepackage[dvips]{graphicx}

\renewcommand{\vec}[1]{\mathbf{#1}}

\newcommand{\eref}[1]{(\ref{#1})}

\renewcommand{\(}{\left(}
\renewcommand{\)}{\right)}

\begin{document}
\title{Spectral scaling laws in MHD turbulence simulations and in the solar wind} 
\author{Stanislav Boldyrev}
\affiliation{Department of Physics, University of Wisconsin-Madison, Madison, WI 53706, USA} 
\author{Jean Carlos Perez}
\affiliation{Department of Physics, University of Wisconsin-Madison, Madison, WI 53706, USA}
\affiliation{Space Science Center and Department of Physics, University of New Hampshire, Durham, NH 03824}  
\author{Joseph E.\ Borovsky} 
\affiliation{Los Alamos National Laboratory, Los Alamos, New Mexico 87545, USA} 
\affiliation{Atmospheric, Oceanic, and Space Sciences Department, University of Michigan, Ann Arbor, MI 48109, USA}
\author{John J.\ Podesta}
\affiliation{Los Alamos National Laboratory, Los Alamos, New Mexico 87545, USA}  

\date{May 18, 2011}

\input psfig.sty

\begin{abstract}
The question is addressed to what extent incompressible magnetohydrodynamics 
(MHD) can describe random magnetic and  velocity fluctuations measured in 
the solar wind. 
It is demonstrated that distributions of spectral indices for the velocity, 
magnetic field, and total energy obtained from high resolution numerical 
simulations are qualitatively and quantitatively similar to solar wind 
observations at 1 AU.  Both simulations and observations show that in the
inertial range the magnetic field spectrum $E_b$ is steeper than 
the velocity spectrum $E_v$ with $E_b \gtrsim E_v$ and that the residual energy
$E_R = E_b-E_v$ decreases nearly following a $k_\perp^{-2}$ scaling.
\end{abstract}

\maketitle

\emph{Introduction.}---Plasma motions in astrophysical systems are usually 
magnetized and turbulent.  At scales larger than characteristic plasma 
kinetic scales, one-fluid magnetohydrodynamics provides a satisfactory 
framework for studying such systems \cite{biskamp03,kulsrud05}. 
Magnetohydrodynamic turbulence has long been invoked to explain the 
properties of the solar wind, where velocity and magnetic field fluctuations 
are measured {\em in situ} over a wide range of scales 
\cite[e.g.,][]{goldstein_rm95,tu_m95,bruno_c05}.  Recent high-resolution 
numerical simulations, however, reported intriguing contradictions with 
the observational data. The Fourier energy spectrum of MHD turbulence 
obtained from numerical simulations appears to have a different scaling 
compared to the scaling inferred from observations.  This raises some 
serious  questions. Do the numerical simulations correctly represent the 
physics of solar wind fluctuations
and, if so, then why doesn't solar wind turbulence exhibit the same 
universal scaling found in 3D MHD simulations?    

To formulate the problem, we rewrite the incompressible MHD equations in terms of the Els\"asser variables, 
\begin{equation}
  \(\frac{\partial}{\partial t}\mp\vec v_A\cdot\nabla\)\vec
  z^\pm+\left(\vec z^\mp\cdot\nabla\right)\vec z^\pm = -\nabla P,
  \label{mhd-elsasser}
\end{equation}
where the Els\"asser variables are defined as $\vec z^\pm=\vec
v\pm\vec b$, $\vec v$ is the fluctuating plasma velocity, $\vec b$ is
the fluctuating magnetic field normalized by $\sqrt{4 \pi \rho_0}$,
${\bf v}_A={\bf B}_0/\sqrt{4\pi \rho_0}$ is the Alfv\'en velocity corresponding to the
uniform magnetic field ${\bf B}_0$, $P=(p/\rho_0+b^2/2)$ includes the
plasma pressure $p$ and the magnetic pressure, $\rho_0$ is the constant 
mass density, and we neglect driving and dissipation terms. The guide field ${\bf B}_0$ can be either imposed 
by external sources or associated with large scale fluctuations that are almost uniform 
with respect to the scales of the inertial range.  It follows from
these equations that for $\vec z^\mp(\vec x,t)=0$, an arbitrary
function $\vec z^\pm(\vec x,t)=F^\pm(\vec x\pm\vec v_At)$ is an exact
nonlinear solution that represents a non-dispersive Alfv\'en  wave propagating
along the direction $\mp\vec v_A$. Nonlinear interactions are thus the result
of collisions between counter-propagating Alfv\'en wave packets.

Denote by $E^\pm=\langle |\vec z^\pm|^2\rangle /4$ the energies associated
with the $\pm$ waves. Those two quantities are independent integrals of motion 
of the ideal MHD system~(\ref{mhd-elsasser}). 
They are related to the total energy and cross-helicity, $E=E^++E^-$ and $H_c=E^+-E^-$, 
respectively. Cross-helicity provides a measure of {\em imbalance} between interacting Alfv\'en modes; when $H_c\neq 0$ the turbulence
is called imbalanced, otherwise it is balanced. The solar wind is essentially imbalanced, as more Alfv\'en waves propagate away from
the sun than toward the sun. In a turbulent state, when energy is supplied to the system at large scales, 
both $E^\pm$ cascade toward small scales where they are damped by viscosity and resistivity. 
Theories and numerical simulations of MHD turbulence address the Fourier 
spectra of the energies $E^{\pm}(k)$ in the inertial range of scales, that is, scales 
much smaller than the forcing scales and much larger than the dissipation scales.  

According to numerical simulations, statistics of MHD turbulence 
are highly anisotropic with respect to the local mean magnetic field. It has recently been argued  
that the field-perpendicular energy spectra of incompressible, homogeneous, strong MHD turbulence 
scale 
as $E^{\pm}({\bf k}_\perp)\propto k_{\perp}^{-3/2}$, in both 
balanced and imbalanced cases \cite{maron_g01,muller_g05,mason_cb06,mason_cb08,perez_b09}, 
and that this scaling is consistent with analytic models \cite{perez_b10}. 
This picture seems however to contradict the observational 
data of the solar wind, which often find the spectrum of {\em magnetic field} fluctuations to be  consistent with the Kolmogorov scaling~$-5/3$ \cite[e.g.,][]{goldstein_rm95}. At the same time, recent measurements of {\em velocity} fluctuations in the solar wind reveal an essentially shallower spectrum, closer to $E_v(k)\propto k^{-3/2}$ \cite[e.g.,][]{podesta_rg07,tessein_etal09,podesta_b10,chen_etal10,borovsky11}. This mismatch in 
the spectral scalings motivated our interest in the problem. To address the problem we notice that there is no strict requirement that the magnetic and 
velocity fluctuations be in equipartition with each other.  Moreover, even though analytic models often appeal to the picture of counter-propagating Alfv\'en modes, with  ${\bf v}= \pm {\bf b}$,  such Alfv\'en modes are 
not statistically independent in strong turbulence. This means, in general, that $\langle {\bf z}^+\cdot {\bf z}^-\rangle = \langle v^2\rangle - \langle b^2\rangle \neq 0$. 

We perform a high-resolution numerical study 
of both balanced and imbalanced MHD turbulence, and concentrate on individual 
magnetic and velocity spectra.  We find that in both cases these spectra are generally not identical. We observe that the so-called residual energy, characterizing the mismatch of the spectra, $E_r(k_\perp)=E_b(k_\perp)-E_v(k_\perp)$, is not universal in that its {\em amplitude} depends of the driving and the degree of imbalance. The {\em scaling} of the residual energy is, however, close to $E_r(k)\propto k_\perp^{-2}$ in both balanced and imbalanced runs. In the balanced case this result was first obtained in  \cite{muller_g05}.
While the total energy spectrum is close to $E(k_\perp)\propto k_\perp^{-3/2}$, the presence of residual energy leads to steeper magnetic spectrum and a shallower   velocity spectrum in an inertial interval of limited extent. However, since the residual spectrum declines faster than the total spectrum, 
the universality of the turbulence should be restored asymptotically at large $k_\perp$, and it can be observed if the inertial interval is large enough. 
 
For a comparison with the solar wind measurements we plot histograms of velocity and magnetic spectral indices measured for individual temporal snapshots in numerical simulations of MHD turbulence. Comparison of the results with analogous histograms obtained from individual solar wind measurements reveals good agreement,  indicating that incompressible MHD provides an adequate framework for modeling MHD-scale turbulence in the solar wind.

\emph{Numerical simulations.}---The universal properties of MHD turbulence are accurately described by
neglecting the parallel component of the fluctuating fields,
associated with the pseudo-Alfv\'en
mode, e.g.,~\cite{galtier_nnp02,galtier_c06,perez_b08}. 
By setting
$\vec z_\|^\pm = 0$ in equation \eref{mhd-elsasser} we obtain the
closed system of equations
\begin{eqnarray}
  \(\frac{\partial}{\partial t}\mp\vec v_A\cdot\nabla_\|\)\vec
  z^\pm+\left(\vec z^\mp\cdot\nabla_\perp\right)\vec z^\pm =
  -\nabla_\perp P\nonumber\\ +\vec f^\pm+\nu\nabla^2\vec z^\pm,
  \label{rmhd-elsasser}
\end{eqnarray}
in which force and dissipation terms have been added to address the
case of steadily driven turbulence, and we assume that viscosity is
equal to resistivity.  This set of equations is known as the Reduced
MHD model (RMHD) \cite{kadomtsev_P74,strauss76}, appropriate for studying 
MHD turbulence with a strong guide field, $v_A\gg v_{rms}$. Numerical 
simulations of full MHD equations show that the universal regime of strong MHD turbulence is reproduced well for $v_A/v_{rms}\geq 5$ \cite[e.g.,][]{muller_bg03,muller_g05,mason_cb08}, which is properly captured by Reduced MHD system~(\ref{rmhd-elsasser}). RMHD allows one to reduce the number of fields by a factor of two and to speed up the numerical integration. We employ a fully dealiased Fourier
pseudo-spectral method to solve equations \eref{rmhd-elsasser} in a rectangular periodic
box, with field-perpendicular cross section $L_\perp^2=(2\pi)^2$ and
field-parallel box size $L_\|=(v_A/v_{rms})L_{\perp}$. 
The choice of a rectangular box, as
discussed in~\cite{perez_b09,perez_b10}, allows for the excitation of
elongated modes at large scales, necessary to avoid a long transition
region between the forcing scale and the beginning of the inertial range. 

The random forcing $\vec f^\pm$ is
applied in Fourier space at wave-numbers $1 \leq k_{\perp} \leq 2,  
k_\| =  2\pi/L_\|$. The Fourier coefficients inside that range
are Gaussian random numbers with amplitude chosen so that the
resulting rms velocity fluctuations are of order unity.  The
individual random values are refreshed independently for each mode on
average every $\tau=0.1~L_\perp/v_{rms}$. We define the Reynolds
number as $Re=(L_\perp/2\pi)v_{rms}/\nu$. In the present simulations, we also introduce correlation between
$\vec v$ and $\vec b$, which is achieved by taking $\vec f^\pm$ as
uncorrelated Gaussian random forces with zero mean and variances $\sigma^2_\pm\equiv \langle |{\bf f}^\pm|^2\rangle$.
 Denoting $ \vec f_v = \frac
12\(\vec f^++\vec f^-\)$, and $\vec f_b = \frac 12\(\vec f^+-\vec f^-\)$,
we obtain that cross-helicity is controlled by $ \langle\vec
f_v\cdot\vec f_b \rangle = \frac 14\(\sigma_+^2-\sigma_-^2\)$. The results presented below 
have been conducted at numerical resolution of $1024^2\times 256$ points, which corresponds to the Reynolds number of $Re\approx 5600$. In the imbalanced run, the normalized cross-helicity is $H_c/E \approx 0.6$.  The cases were run for up to 200 large-scale eddy turnover times in order to get reliable statistics, see \cite{perez_b10}. 

\begin{figure} [tbp]\label{spectra}
\centerline{\psfig{file=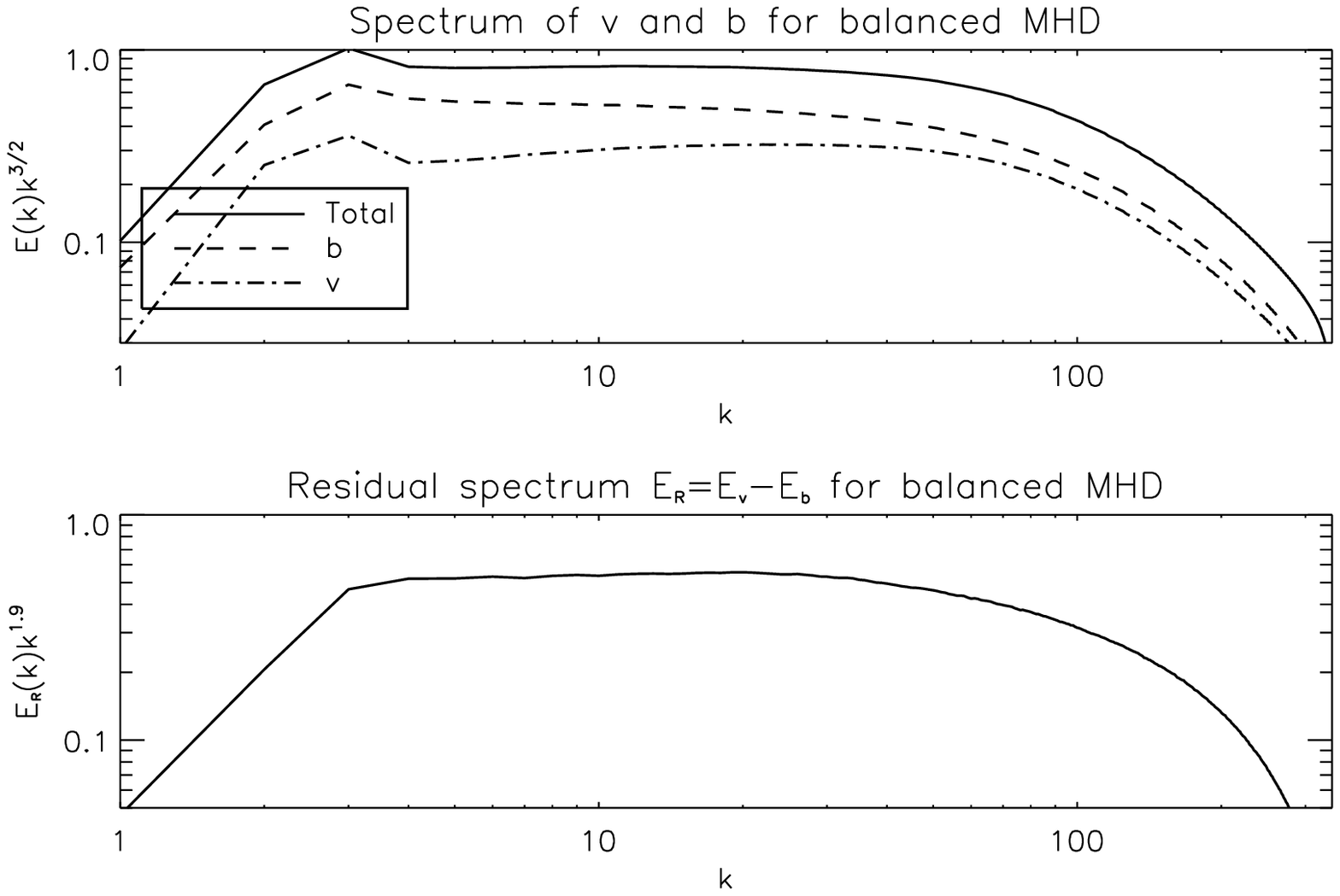,width=3.6in,angle=0}}
\centerline{\psfig{file=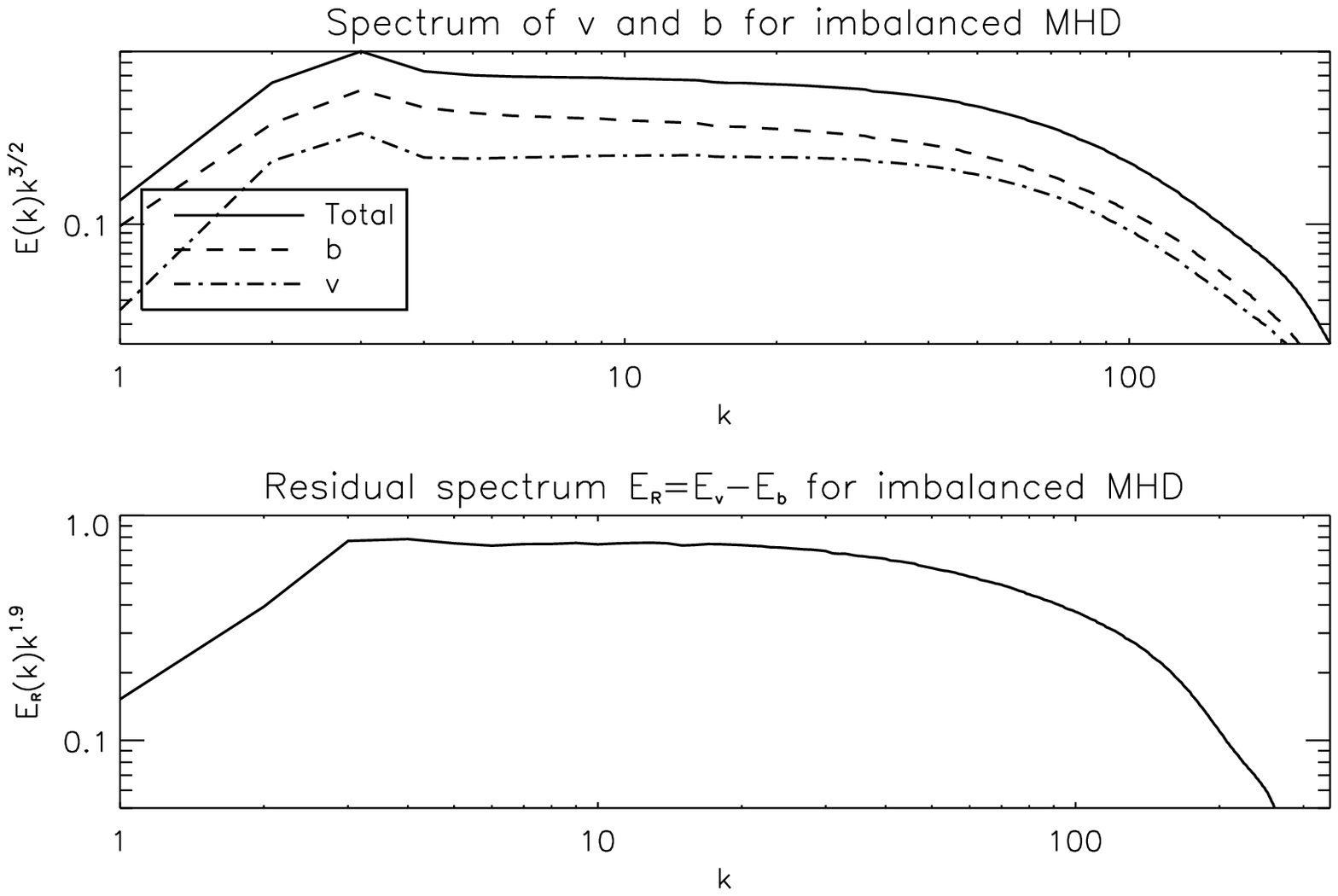,width=3.6in,angle=0}}
\caption{Spectra of kinetic, magnetic, total, and residual energies in numerical simulations of balanced and imbalanced MHD turbulence.}
\end{figure}


The results of numerical simulations are presented in Figs.~1 and 2. 
Two important observations should be made here. First, there is a tendency 
of magnetic energy to exceed the kinetic energy at large inertial range 
scales for both balanced and imbalanced cases. The presence of nonzero residual energy was noted in previous studies \cite[e.g.,][]{pouquet_fl76,muller_g05}. Comparison of our results with previously available numerical data suggests that the level of residual energy is not universal, rather, it can be affected by the driving and the degree of imbalance. 
Second, the excess of magnetic energy persists in the whole inertial interval, however in quite a peculiar fashion. In both balanced and imbalanced cases, the residual energy spectrum has a power-law behavior close to $E_r(k_\perp)\propto k_\perp^{-2}$.  In an inertial interval of limited extent, this leads to steepening of the magnetic spectrum and flattening of the velocity spectrum, however, the total energy scaling  stays close to $-3/2$. Due to the relatively rapid spectral decline, the residual energy provides a subdominant contribution to both kinetic and magnetic energy spectra. We therefore propose that the mismatch between velocity and magnetic field energies becomes  asymptotically irrelevant as the inertial range increases, in which case the universal scaling $-3/2$ is restored for both $E_v(k_\perp)$ and $E_b(k_\perp)$.

\emph{Comparison with solar wind data.}---
Comparison of available numerical simulations with solar wind data is 
complicated by the fact that individual solar wind measurements typically 
last for a few correlation times ($\tau_c \sim$1 hour), while in numerical 
simulations the spectra are averaged over tens or hundreds of turnover 
times to obtain good convergence. To make an appropriate comparison, we 
measure the velocity and magnetic field spectra for many individual 
simulation snapshots separated by about one eddy turnover time. We then 
plot distributions of the spectral indices obtained in this way for both 
balanced and imbalanced cases. The results are presented in Fig.~(2).  

\begin{figure}[tbp]
\centerline{\psfig{file=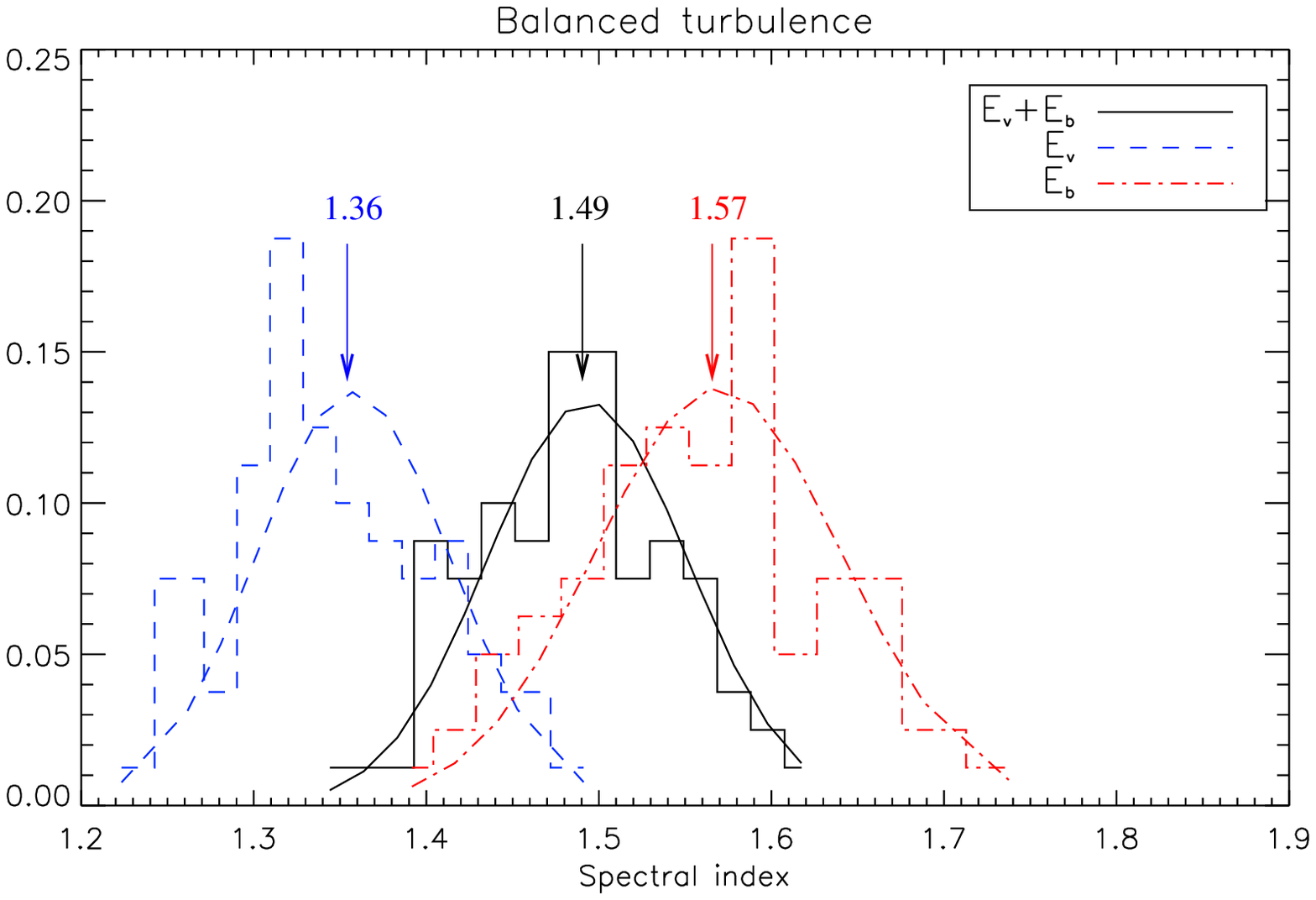,width=3.6in,angle=0}}
\centerline{\psfig{file=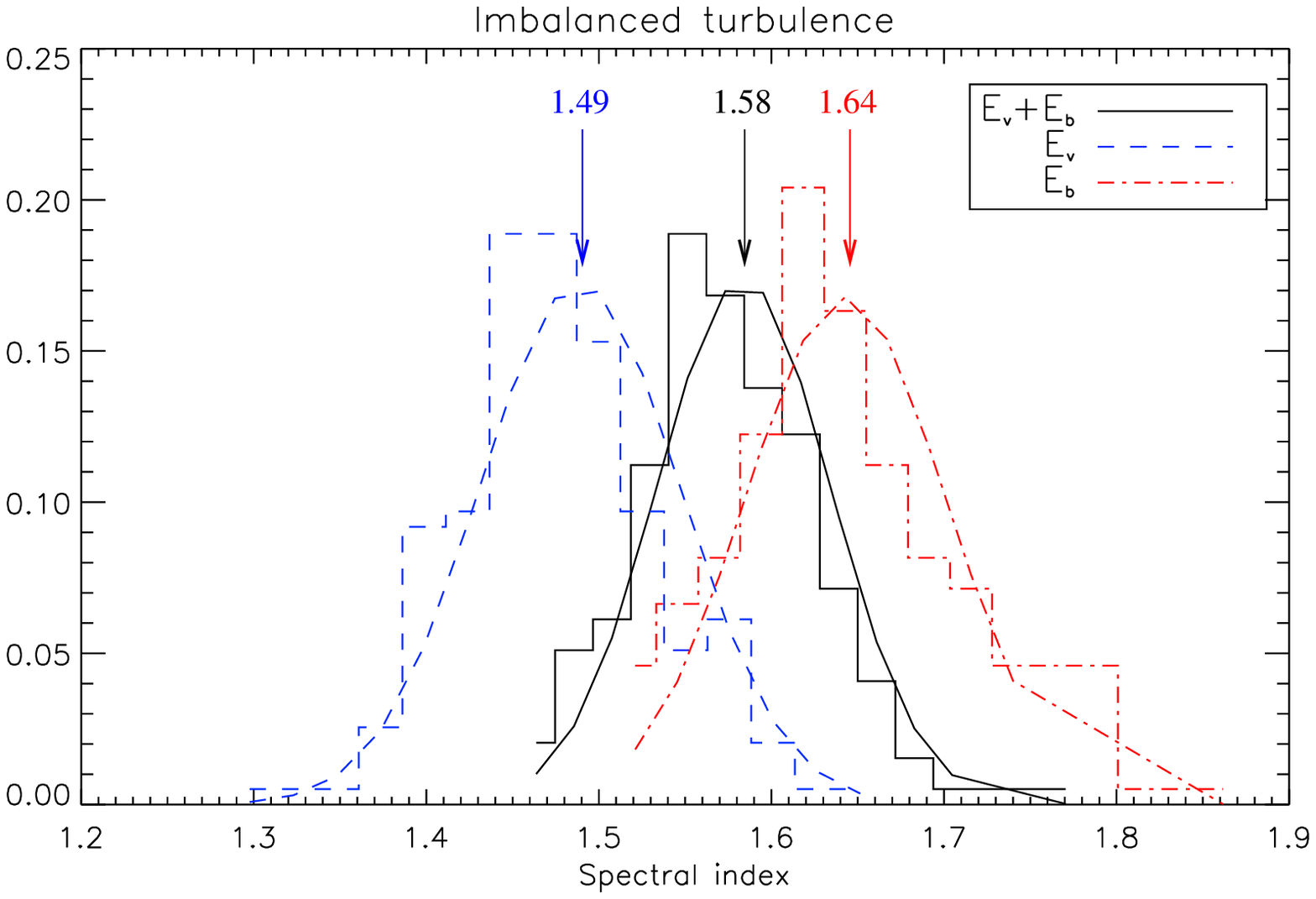,width=3.6in,angle=0}}
\label{hist}
\caption{Distributions of spectral indices for kinetic, magnetic, and total energies for individual snapshots in numerical simulations of MHD turbulence. Upper plot: balanced turbulence, 80 snapshots; lower plot: imbalanced turbulence, 196 snapshots. The average spectral indices are indicated by arrows. Normal distributions with the mean values and variances matching those of the data are also shown.}
\end{figure}

\begin{figure}[!t]
    \includegraphics[width=\columnwidth]{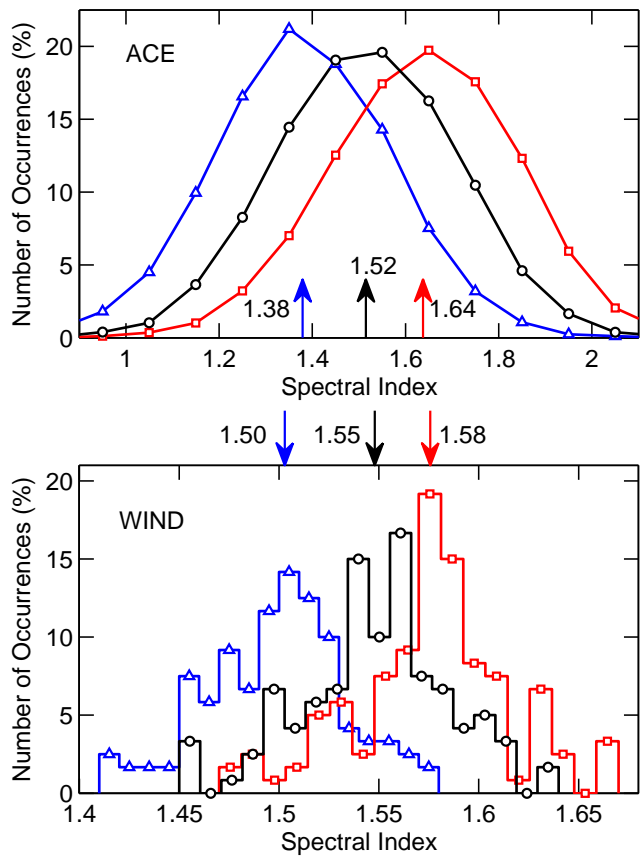}
    \caption{Histograms of measured spectral indices for the velocity spectrum (blue triangles),
magnetic field spectrum (red squares), and total energy spectrum (black circles)
in the solar wind using data from the {\it ACE} and {\it Wind} spacecraft.   The average
spectral indices are indicated by the arrows.  Note the different horizontal scales in the two plots.}
    \label{solar_hist}
  \end{figure}

For comparison, Fig.~(\ref{solar_hist}) presents analogous histograms obtained 
using solar wind measurements. The data from the {\it Advanced Composition 
Explorer} (ACE) spacecraft consists of 15,472 different spectra covering 
the 10 year period from 1998 to 2008 \cite{borovsky11}.  The data from the 
{\it Wind} spacecraft consists of those 120 of 176 spectra studied in
\cite{podesta_b10} having the highest normalized cross-helicity 
$|\sigma_c|>0.76$, that is, the greatest imbalance.  The spectral indices are obtained from fits over the range of
spacecraft frame frequencies from $10^{-3}$ to $3\times 10^{-2}$~Hz for the Wind data \cite{podesta_b10}
and from $1.8\times 10^{-4}$ to $3.9\times 10^{-3}$~Hz for the ACE data \cite{borovsky11}. It turns out that the 
scatter of individual indices in numerical simulations closely resembles 
the corresponding scatter in solar wind measurements. The excess of 
magnetic energy over kinetic energy is also seen in the solar wind where,
in the study \cite{podesta_b10} for example, 
the power-law exponent of the residual energy takes typical values around 
1.75. The solar wind also shows a tendency for the spectral indices of velocity 
and magnetic field to be closer together when the normalized cross-helicity is
high than when it is low \cite{borovsky_d10,podesta_b10}.  A similar tendency is evident 
in the simulation results in Fig.~2. 
We therefore propose that the mismatch between $E_b(k_\perp)$ and 
$E_v(k_\perp)$ observed in the solar wind turbulence is neither the 
manifestation of non-universality of MHD turbulence nor does it indicate 
a breakdown of the applicability of incompressible MHD turbulence theory to 
the solar wind. Rather, it is a consequence of 
significant residual energy generated at large scales, 
in  agreement with numerical simulations.


\emph{Conclusions.}---Velocity spectra, magnetic field spectra, and total 
energy spectra in high resolution numerical simulations of 3D  
incompressible MHD turbulence are shown to be in good agreement with solar wind 
observations at 1 AU where the respective spectral indicies of $E_v$, $E_b$, 
and $E$ are approximately centered around 1.4-1.5, 1.6-1.7, and 1.5-1.6 
\cite[see, e.g.,][]{tessein_etal09,podesta_b10,borovsky11}.  It is important 
to note that the large variability found in solar wind spectral 
indices is also observed in temporal snapshots of the numerical 
simulations. The unique scaling laws obtained by
averaging simulation spectra over many eddy turnover times can also 
be obtained through a statistical analysis of the spectral indices of those
snapshots.  This provides justification for the widely used
statistical approach to the analysis of spectral indices in the solar wind, where 
averaging over many turnover times is not practical. Our results indicate 
that  universal inertial range dynamics may be present in the solar wind 
in spite of the observed high variability of solar wind measurements, and 
that solar wind turbulence spectra are consistent with the characteristics of  
incompressible MHD turbulence.

\acknowledgments This work was supported   
by the US DoE awards DE-FG02-07ER54932, DE-SC0003888, DE-SC0001794, the NSF grant PHY-0903872, and the NSF Center for magnetic Self-organization in Laboratory and Astrophysical Plasmas at U. Wisconsin-Madison.  
High Performance Computing resources were
provided by the Texas Advanced Computing Center (TACC) at the
University of Texas at Austin under the NSF-Teragrid Project
TG-PHY080013N. Work at Los Alamos was supported by the NASA Solar and 
Heliospheric Physics Program and the NSF SHINE Program.


\begin{thebibliography}{24}
\expandafter\ifx\csname natexlab\endcsname\relax\def\natexlab#1{#1}\fi
\expandafter\ifx\csname bibnamefont\endcsname\relax
  \def\bibnamefont#1{#1}\fi
\expandafter\ifx\csname bibfnamefont\endcsname\relax
  \def\bibfnamefont#1{#1}\fi
\expandafter\ifx\csname citenamefont\endcsname\relax
  \def\citenamefont#1{#1}\fi
\expandafter\ifx\csname url\endcsname\relax
  \def\url#1{\texttt{#1}}\fi
\expandafter\ifx\csname urlprefix\endcsname\relax\def\urlprefix{URL }\fi
\providecommand{\bibinfo}[2]{#2}
\providecommand{\eprint}[2][]{\url{#2}}

\bibitem[{\citenamefont{{Biskamp}}(2003)}]{biskamp03}
\bibinfo{author}{\bibfnamefont{D.}~\bibnamefont{{Biskamp}}},
  \emph{\bibinfo{title}{{Magnetohydrodynamic Turbulence}}}
  (\bibinfo{year}{2003}).

\bibitem[{\citenamefont{{Kulsrud}}(2005)}]{kulsrud05}
\bibinfo{author}{\bibfnamefont{R.~M.} \bibnamefont{{Kulsrud}}},
  \emph{\bibinfo{title}{{Plasma physics for astrophysics}}}
  (\bibinfo{year}{2005}).

\bibitem[{\citenamefont{{Goldstein} et~al.}(1995)\citenamefont{{Goldstein},
  {Roberts}, and {Matthaeus}}}]{goldstein_rm95}
\bibinfo{author}{\bibfnamefont{M.~L.} \bibnamefont{{Goldstein}}},
  \bibinfo{author}{\bibfnamefont{D.~A.} \bibnamefont{{Roberts}}},
  \bibnamefont{and} \bibinfo{author}{\bibfnamefont{W.~H.}
  \bibnamefont{{Matthaeus}}}, \bibinfo{journal}{Ann. Rev. Astron. Astrophys.}
  \textbf{\bibinfo{volume}{33}}, \bibinfo{pages}{283} (\bibinfo{year}{1995}).

\bibitem[{\citenamefont{{Tu} and {Marsch}}(1995)}]{tu_m95}
\bibinfo{author}{\bibfnamefont{C.}~\bibnamefont{{Tu}}} \bibnamefont{and}
  \bibinfo{author}{\bibfnamefont{E.}~\bibnamefont{{Marsch}}},
  \bibinfo{journal}{Space Sc. Rev.} \textbf{\bibinfo{volume}{73}},
  \bibinfo{pages}{1} (\bibinfo{year}{1995}).

\bibitem[{\citenamefont{{Bruno} and {Carbone}}(2005)}]{bruno_c05}
\bibinfo{author}{\bibfnamefont{R.}~\bibnamefont{{Bruno}}} \bibnamefont{and}
  \bibinfo{author}{\bibfnamefont{V.}~\bibnamefont{{Carbone}}},
  \bibinfo{journal}{Living Reviews in Solar Physics}
  \textbf{\bibinfo{volume}{2}}, \bibinfo{pages}{4} (\bibinfo{year}{2005}).

\bibitem[{\citenamefont{{Maron} and {Goldreich}}(2001)}]{maron_g01}
\bibinfo{author}{\bibfnamefont{J.}~\bibnamefont{{Maron}}} \bibnamefont{and}
  \bibinfo{author}{\bibfnamefont{P.}~\bibnamefont{{Goldreich}}},
  \bibinfo{journal}{Astrophys. J.} \textbf{\bibinfo{volume}{554}},
  \bibinfo{pages}{1175} (\bibinfo{year}{2001}),
  \eprint{arXiv:astro-ph/0012491}.

\bibitem[{\citenamefont{{M{\"u}ller} and {Grappin}}(2005)}]{muller_g05}
\bibinfo{author}{\bibfnamefont{W.}~\bibnamefont{{M{\"u}ller}}}
  \bibnamefont{and}
  \bibinfo{author}{\bibfnamefont{R.}~\bibnamefont{{Grappin}}},
  \bibinfo{journal}{Physical Review Letters} \textbf{\bibinfo{volume}{95}},
  \bibinfo{pages}{114502} (\bibinfo{year}{2005}),
  \eprint{arXiv:physics/0509019}.

\bibitem[{\citenamefont{{Mason} et~al.}(2006)\citenamefont{{Mason}, {Cattaneo},
  and {Boldyrev}}}]{mason_cb06}
\bibinfo{author}{\bibfnamefont{J.}~\bibnamefont{{Mason}}},
  \bibinfo{author}{\bibfnamefont{F.}~\bibnamefont{{Cattaneo}}},
  \bibnamefont{and}
  \bibinfo{author}{\bibfnamefont{S.}~\bibnamefont{{Boldyrev}}},
  \bibinfo{journal}{Physical Review Letters} \textbf{\bibinfo{volume}{97}},
  \bibinfo{pages}{255002} (\bibinfo{year}{2006}),
  \eprint{arXiv:astro-ph/0602382}.

\bibitem[{\citenamefont{{Mason} et~al.}(2008)\citenamefont{{Mason}, {Cattaneo},
  and {Boldyrev}}}]{mason_cb08}
\bibinfo{author}{\bibfnamefont{J.}~\bibnamefont{{Mason}}},
  \bibinfo{author}{\bibfnamefont{F.}~\bibnamefont{{Cattaneo}}},
  \bibnamefont{and}
  \bibinfo{author}{\bibfnamefont{S.}~\bibnamefont{{Boldyrev}}},
  \bibinfo{journal}{Physical Review E} \textbf{\bibinfo{volume}{77}},
  \bibinfo{pages}{036403} (\bibinfo{year}{2008}), \eprint{0706.2003}.

\bibitem[{\citenamefont{{Perez} and {Boldyrev}}(2009)}]{perez_b09}
\bibinfo{author}{\bibfnamefont{J.~C.} \bibnamefont{{Perez}}} \bibnamefont{and}
  \bibinfo{author}{\bibfnamefont{S.}~\bibnamefont{{Boldyrev}}},
  \bibinfo{journal}{Physical Review Letters} \textbf{\bibinfo{volume}{102}},
  \bibinfo{pages}{025003} (\bibinfo{year}{2009}), \eprint{0807.2635}.

\bibitem[{\citenamefont{{Perez} and {Boldyrev}}(2010)}]{perez_b10}
\bibinfo{author}{\bibfnamefont{J.~C.} \bibnamefont{{Perez}}} \bibnamefont{and}
  \bibinfo{author}{\bibfnamefont{S.}~\bibnamefont{{Boldyrev}}},
  \bibinfo{journal}{Astrophys. J. Lett.} \textbf{\bibinfo{volume}{710}},
  \bibinfo{pages}{L63} (\bibinfo{year}{2010}), \eprint{0912.0901}.

\bibitem[{\citenamefont{{Podesta} et~al.}(2007)\citenamefont{{Podesta},
  {Roberts}, and {Goldstein}}}]{podesta_rg07}
\bibinfo{author}{\bibfnamefont{J.~J.} \bibnamefont{{Podesta}}},
  \bibinfo{author}{\bibfnamefont{D.~A.} \bibnamefont{{Roberts}}},
  \bibnamefont{and} \bibinfo{author}{\bibfnamefont{M.~L.}
  \bibnamefont{{Goldstein}}}, \bibinfo{journal}{Astrophys. J.}
  \textbf{\bibinfo{volume}{664}}, \bibinfo{pages}{543} (\bibinfo{year}{2007}).

\bibitem[{\citenamefont{{Tessein} et~al.}(2009)\citenamefont{{Tessein},
  {Smith}, {MacBride}, {Matthaeus}, {Forman}, and {Borovsky}}}]{tessein_etal09}
\bibinfo{author}{\bibfnamefont{J.~A.} \bibnamefont{{Tessein}}},
  \bibinfo{author}{\bibfnamefont{C.~W.} \bibnamefont{{Smith}}},
  \bibinfo{author}{\bibfnamefont{B.~T.} \bibnamefont{{MacBride}}},
  \bibinfo{author}{\bibfnamefont{W.~H.} \bibnamefont{{Matthaeus}}},
  \bibinfo{author}{\bibfnamefont{M.~A.} \bibnamefont{{Forman}}},
  \bibnamefont{and} \bibinfo{author}{\bibfnamefont{J.~E.}
  \bibnamefont{{Borovsky}}}, \bibinfo{journal}{Astrophys. J.}
  \textbf{\bibinfo{volume}{692}}, \bibinfo{pages}{684} (\bibinfo{year}{2009}).

\bibitem[{\citenamefont{{Podesta} and {Borovsky}}(2010)}]{podesta_b10}
\bibinfo{author}{\bibfnamefont{J.~J.} \bibnamefont{{Podesta}}}
  \bibnamefont{and} \bibinfo{author}{\bibfnamefont{J.~E.}
  \bibnamefont{{Borovsky}}}, \bibinfo{journal}{Physics of Plasmas}
  \textbf{\bibinfo{volume}{17}}, \bibinfo{pages}{112905}
  (\bibinfo{year}{2010}).

\bibitem[{\citenamefont{{Chen} et~al.}(2010)\citenamefont{{Chen}, {Mallet},
  {Yousef}, {Schekochihin}, and {Horbury}}}]{chen_etal10}
\bibinfo{author}{\bibfnamefont{C.~H.~K.} \bibnamefont{{Chen}}},
  \bibinfo{author}{\bibfnamefont{A.}~\bibnamefont{{Mallet}}},
  \bibinfo{author}{\bibfnamefont{T.~A.} \bibnamefont{{Yousef}}},
  \bibinfo{author}{\bibfnamefont{A.~A.} \bibnamefont{{Schekochihin}}},
  \bibnamefont{and} \bibinfo{author}{\bibfnamefont{T.~S.}
  \bibnamefont{{Horbury}}}, \bibinfo{journal}{ArXiv e-prints}
  (\bibinfo{year}{2010}), \eprint{1009.0662}.

\bibitem[{\citenamefont{{Borovsky}}(2011)}]{borovsky11}
\bibinfo{author}{\bibfnamefont{J.~E.} \bibnamefont{{Borovsky}}},
  \bibinfo{journal}{submitted to J Geophys. Res.}  (\bibinfo{year}{2011}).

\bibitem[{\citenamefont{{Galtier} et~al.}(2002)\citenamefont{{Galtier},
  {Nazarenko}, {Newell}, and {Pouquet}}}]{galtier_nnp02}
\bibinfo{author}{\bibfnamefont{S.}~\bibnamefont{{Galtier}}},
  \bibinfo{author}{\bibfnamefont{S.~V.} \bibnamefont{{Nazarenko}}},
  \bibinfo{author}{\bibfnamefont{A.~C.} \bibnamefont{{Newell}}},
  \bibnamefont{and}
  \bibinfo{author}{\bibfnamefont{A.}~\bibnamefont{{Pouquet}}},
  \bibinfo{journal}{Astrophys. J. Lett.} \textbf{\bibinfo{volume}{564}},
  \bibinfo{pages}{L49} (\bibinfo{year}{2002}).

\bibitem[{\citenamefont{{Galtier} and {Chandran}}(2006)}]{galtier_c06}
\bibinfo{author}{\bibfnamefont{S.}~\bibnamefont{{Galtier}}} \bibnamefont{and}
  \bibinfo{author}{\bibfnamefont{B.~D.~G.} \bibnamefont{{Chandran}}},
  \bibinfo{journal}{Physics of Plasmas} \textbf{\bibinfo{volume}{13}},
  \bibinfo{pages}{114505} (\bibinfo{year}{2006}).

\bibitem[{\citenamefont{{Perez} and {Boldyrev}}(2008)}]{perez_b08}
\bibinfo{author}{\bibfnamefont{J.~C.} \bibnamefont{{Perez}}} \bibnamefont{and}
  \bibinfo{author}{\bibfnamefont{S.}~\bibnamefont{{Boldyrev}}},
  \bibinfo{journal}{Astrophys. J. Lett.} \textbf{\bibinfo{volume}{672}},
  \bibinfo{pages}{L61} (\bibinfo{year}{2008}), \eprint{0712.2086}.

\bibitem[{\citenamefont{{Kadomtsev} and {Pogutse}}(1974)}]{kadomtsev_P74}
\bibinfo{author}{\bibfnamefont{B.~B.} \bibnamefont{{Kadomtsev}}}
  \bibnamefont{and} \bibinfo{author}{\bibfnamefont{O.~P.}
  \bibnamefont{{Pogutse}}}, \bibinfo{journal}{Soviet Journal of Experimental
  and Theoretical Physics} \textbf{\bibinfo{volume}{38}}, \bibinfo{pages}{283}
  (\bibinfo{year}{1974}).

\bibitem[{\citenamefont{{Strauss}}(1976)}]{strauss76}
\bibinfo{author}{\bibfnamefont{H.~R.} \bibnamefont{{Strauss}}},
  \bibinfo{journal}{Physics of Fluids} \textbf{\bibinfo{volume}{19}},
  \bibinfo{pages}{134} (\bibinfo{year}{1976}).

\bibitem[{\citenamefont{{M{\"u}ller} et~al.}(2003)\citenamefont{{M{\"u}ller},
  {Biskamp}, and {Grappin}}}]{muller_bg03}
\bibinfo{author}{\bibfnamefont{W.}~\bibnamefont{{M{\"u}ller}}},
  \bibinfo{author}{\bibfnamefont{D.}~\bibnamefont{{Biskamp}}},
  \bibnamefont{and}
  \bibinfo{author}{\bibfnamefont{R.}~\bibnamefont{{Grappin}}},
  \bibinfo{journal}{Physical Review E} \textbf{\bibinfo{volume}{67}},
  \bibinfo{pages}{066302} (\bibinfo{year}{2003}),
  \eprint{arXiv:physics/0306045}.

\bibitem[{\citenamefont{{Pouquet} et~al.}(1976)\citenamefont{{Pouquet},
  {Frisch}, and {Leorat}}}]{pouquet_fl76}
\bibinfo{author}{\bibfnamefont{A.}~\bibnamefont{{Pouquet}}},
  \bibinfo{author}{\bibfnamefont{U.}~\bibnamefont{{Frisch}}}, \bibnamefont{and}
  \bibinfo{author}{\bibfnamefont{J.}~\bibnamefont{{Leorat}}},
  \bibinfo{journal}{Journal of Fluid Mechanics} \textbf{\bibinfo{volume}{77}},
  \bibinfo{pages}{321} (\bibinfo{year}{1976}).

\bibitem[{\citenamefont{{Borovsky} and {Denton}}(2010)}]{borovsky_d10}
\bibinfo{author}{\bibfnamefont{J.~E.} \bibnamefont{{Borovsky}}}
  \bibnamefont{and} \bibinfo{author}{\bibfnamefont{M.~H.}
  \bibnamefont{{Denton}}}, \bibinfo{journal}{Journal of Geophysical Research
  (Space Physics)} \textbf{\bibinfo{volume}{115}}, \bibinfo{pages}{10101}
  (\bibinfo{year}{2010}).

\end{thebibliography}

\end {document}